Title: Why understanding multiplex social network structuring processes will help us better understand the evolution of human behavior

Running title: Multiplex networks and behavior

Authors: Curtis Atkisson1, Piotr J. Górski2, Matthew O. Jackson3,4,5, Janusz A. Hołyst2,6, Raissa M. D'Souza7,4

1: Department of Anthropology, University of California, Davis
2: Faculty of Physics, Warsaw University of Technology
3: Department of Economics, Stanford University
4: External faculty, Santa Fe Institute
5: Fellow of CIFAR
6: ITMO University
7: Department of Computer Science, Department of Mechanical and Aerospace Engineering, University of California, Davis

Corresponding author: Curtis Atkisson, atkissoncj@gmail.com

**ABSTRACT**

Social scientists have long appreciated that relationships between individuals cannot be described from observing a single domain, and that the structure across domains of interaction can have important effects on outcomes of interest (e.g., cooperation).1 One debate explicitly about this surrounds food sharing. Some argue that failing to find reciprocal food sharing means that some process other than reciprocity must be occurring, whereas others argue for models that allow reciprocity to span domains in the form of trade.2 Multilayer networks, high-dimensional networks that allow us to consider multiple sets of relationships at the same time, are ubiquitous and have consequences, so processes giving rise to them are important social phenomena. The analysis of multi-dimensional social networks has recently garnered the attention of the network science community.3 Recent models of these processes show how ignoring layer interdependencies can lead one to miss why a layer formed the way it did, and/or draw erroneous conclusions.6 Understanding the structuring processes that underlie multiplex networks will help understand increasingly rich datasets, giving more accurate and complete pictures of social interactions.

**1 Social networks, multilayer networks, and multiplex networks**

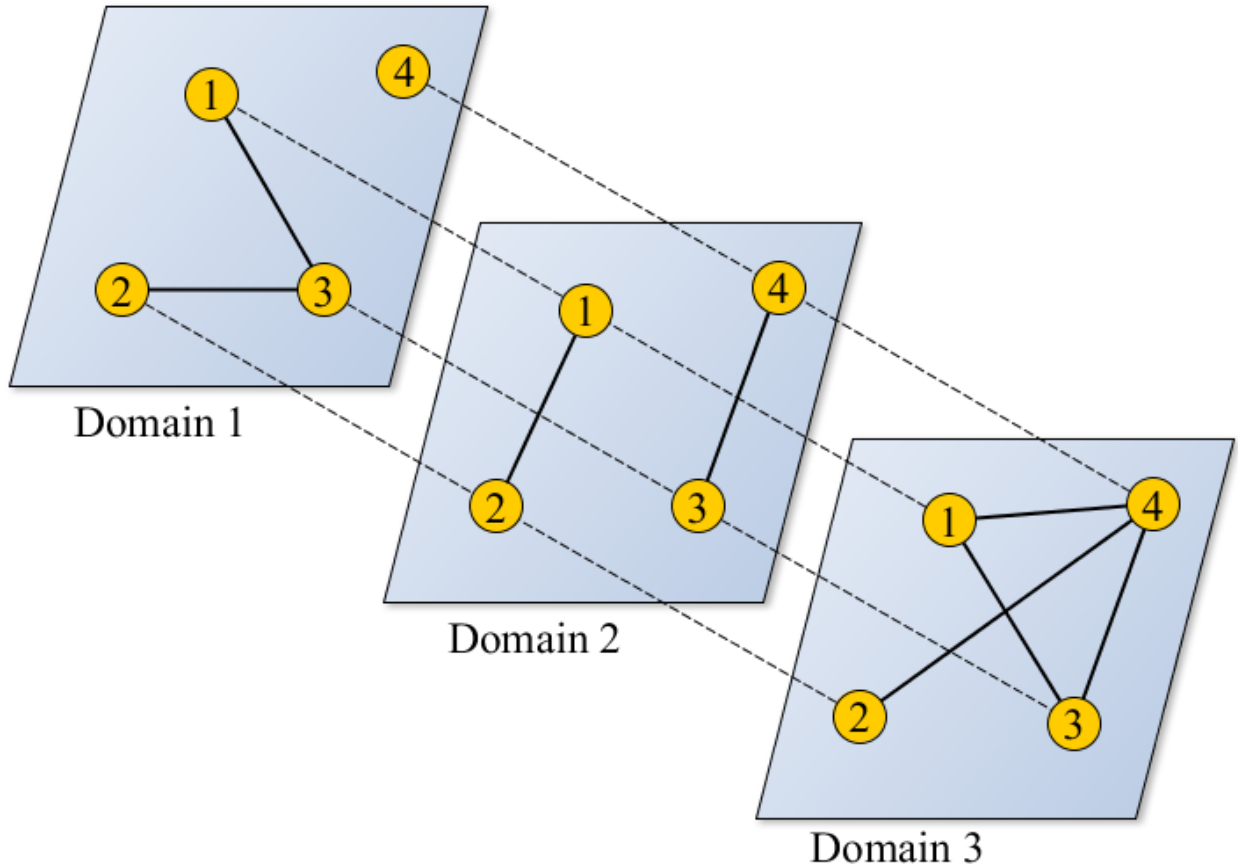

Fig 1: Multiplex networks. The set of individuals is the same across all layers. Individuals are not connected to others across layers but can be connected to different sets of people on different layers.

Social networks are representations of relationships that allow us to use methods from graph theory.4 Networks consist of nodes, which may be represented as individuals, connected to each other by ties. The category of multilayer networks encompasses all networks consisting of more than one set of nodes and/or ties, where each layer is defined as a unique set of nodes and ties. Multiplex networks are the subset of multilayer networks with two basic properties: (1) all layers share the same set of nodes (i.e., each node replicated in each layer) and (2) all nodes are connected only to themselves across layers (see Fig 1). One example of a multiplex network is a social network with layers formed by different domains of interactions, such as hunting, farming, and drinking. In such a domain-specific multiplex network, all individuals could do all those things (i.e., the same set of nodes is shared across domains), but they may do different things with different sets of people.8

**2 Network structuring processes**

We can consider a benchmark model with no constraints. Without costs or interdependencies, individuals would optimize each of their networks by rearranging their relationships. Individuals, however, may be unable to do this due to features of the existing network itself or other reasons, e.g. time constraints. We call the rules for how a network changes based on the current features of the networks and the individuals that compose them network structuring processes. These conditions affect the likelihood of a tie arising between two individuals in a given domain or change individuals' network-based outcomes due to their pattern of ties.

We briefly highlight a few network structuring processes that arise in the context of multiplex networks. Ties might arise in multiple domains between the same individuals because features of the individuals that make a tie likely in one domain are also operating in other domains. This may include things like personality or risk tolerance: individuals who are wary of being caught alone after dark may fish together in mid-day and chop firewood in the evening, also together. Ties between individuals in one domain may be more likely if they are connected in other domains. Examples are incidental network membership (discussed in detail below), as well as benefits to bundling relationships: a person who is a great hunter but poor fisher in an environment of high day-to-day variability in domain-specific returns might offer to be an exchange partner in both domains with someone who is a poor hunter but great fisher. Individuals may struggle to reorganize their networks if the probability of removing a tie depends on other domains of the network. This includes such processes as constraining outside options: the excellent hunter might threaten to not hunt with the excellent fisher if the excellent fisher does not fish with them. Finally, outcomes may be the result of interactions between domains. This includes processes such as alignment of incentives: if a hungry hunter is a poor hunter in a cooperative hunting exercise, then that individual's partners in the hunting domain might share additional food with him, therefore having a connection in the food sharing domain, so that hunting returns are higher for everyone.

**3 Incidental network membership**

We now discuss one important but specific example of a multiplex network structuring process to illustrate some of our main points. The process of incidental network membership rests on a few key premises. First, relationships require time and effort. Second, organisms do not have infinite time and resources. Third, relationships in some domains have a higher net benefit. If these premises are true, then organisms will prioritize optimizing networks in the domains with the highest net benefit. Given finite time and resources, organisms may optimize their entire multiplex network by extending a relationship with a partner on one important domain into a less important domain—even if that individual is not an optimal partner in the other domain. This can result in non-optimal networks when considered as a single layer.

As an illustration, the Makushi of southern Guyana grow and process cassava into a product that is shelf-stable for years by parching the cassava with beef fat to remove the water to make what they call farine.5 Processing cassava to make farine involves many steps, which must occur concurrently. Because of this, it is the best use of time to have several people working together on different stages of the process, constantly adding more cassava to the farine pan. Indeed, women (who do most of this work) have preferred cassava parching partners and are rather consistent in their use of those partners (CA observation). These women spend large amounts of time together, talking constantly. It is common to hear women seeking out advice on their personal lives or reproductive decisions. Since these women have already received such information as a by-product of their cassava processing, they may not be motivated to pay an additional cost to recruit better partners in their advice network for only marginally better information. By increasing the efficiency of one layer of the multiplex network (cassava processing), inefficiencies have been introduced on another layer (the reproductive advice network).

**4 A model of a network structuring process**

We now discuss a formal model of a network structuring process by Górski and colleagues in more detail.6 This model examines how coupling between two layers of a multiplex network impacts the probability of reaching a system equilibrium (a type of network optimality), such that, looking from single-layered perspective, all individuals are happy with the relationships they have. In the real world, this "network optimality" implies that individuals are able to get and maintain the relationships they would like. In this model, each tie in each domain possesses a real weight ranging from -1 to +1. Positive and negative weights correspond to good and bad relations between individuals, respectively. The weights can change in time and the change of a tie weight between two individuals in a domain at each time step is determined by their relationship at the previous time step, their relationships with neighbors they share in common in the focal domain, and the current tie weight in the other domain. The impact of the current tie weight in the other domain can vary in strength due to coupling between layers. The coupling between layers in this model can be asymmetrical such that a tie weight in layer A changes more in response to the tie weight in layer B than the reverse. An example of this would be that people already processing cassava together can give each other reproductive advice since they are spending the time together anyway, but those already giving reproductive advice may not have cassava around to process together. The analysis of the model finds that if layers are completely disconnected from each other, network optimality is achieved independently in each of the domains.6 If one layer is much more strongly driven by the other layer than the reverse, network optimality is achieved because the dominant layer will drag the other layer to its state. But for the parameter space between those extremes, where both layers impact each other, network optimality may not always be achieved. Furthermore, the parameter space in the coupling strength for which optimality is often achieved decreases as network size increases. Figure 2 shows the probability of optimality given different coupling strengths for networks of four sizes. The results for each network size is given in its own panel. This figure, therefore, allows us to compare the effect of coupling strength on the probability of reaching optimality for four networks of different sizes. This model demonstrates that ignoring the coupled nature of networks may result in failing to find expected relationships between networks and outcomes. This conclusion may be unwarranted, however, because the underlying structure was not accounted for. We give some thoughts about how this may be addressed in data below.

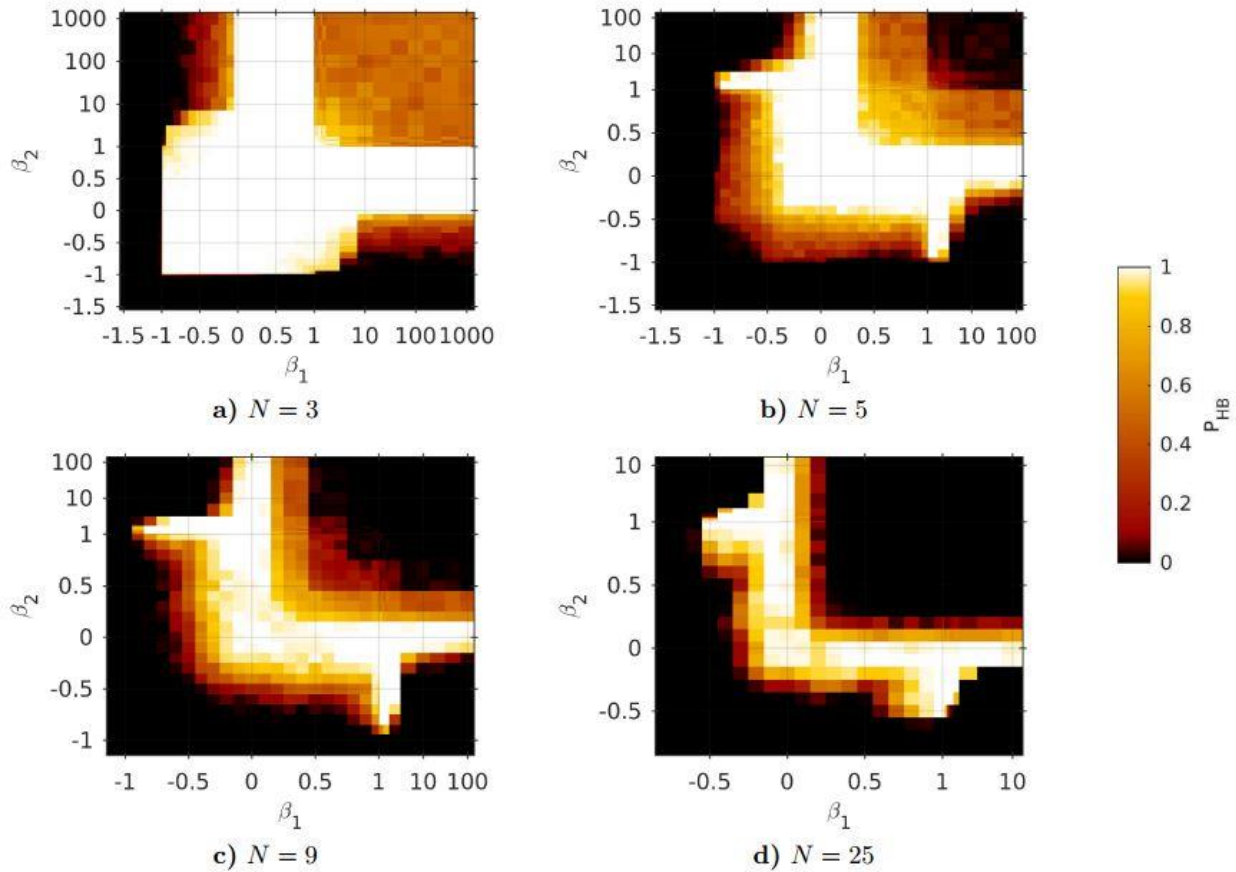

Fig 2: Observing only a single layer of a multiplex network may lead a researcher to wrong conclusions. This figure shows the probability of networks of size N with randomly generated initial tie weights and two layers reaching optimality for the multiplex (O).6 β1 and β2 are coupling coefficients. For instance, β1 represents the influence of the tie weights in layer 2 on the tie weights in layer 1, and β2 the reverse. The color at each pixel shows O, for that combination of β1 and β2, ranging from black (O=0) to white (O=1). An uncoupled layer (β1=β2=0) always reaches optimal states (O=1). However, the coupling between layers decreases the probability of reaching optimality. Therefore, a researcher ignoring the other layer may draw wrong conclusions about individual layers in the multiplex–the researcher might wrongly think to have discovered non-optimality in the single-layered network.

This model leads us to a central formal finding in the nascent study of multiplex network structuring processes: looking at domain-specific networks without appreciating the multiplex structure can lead us to the wrong conclusions (e.g., assuming each layer of the network rather than the entire network is being optimized). This applies to research that concentrates solely on network structure and/or formation, as well as research studying the effect of networks on outcomes. If we examine a single domain to find optimality, we are unlikely to find optimality simply because we have not examined the whole network: agents will be optimizing across their entire multiplex network.

**5 Analyzing social networks given multiplexity**

Models of multiplex networks indicate that their effects are pervasive. The development of tools to analyze these types of networks has thus far proceeded by network scientists using publicly available datasets such as transportation networks. Many anthropologists spend extensive amounts of time observing how complex behaviors are enacted in small-scale societies. The small scale of these networks provides a tractable dataset from which one may draw insights. Because of this, the insights of anthropologists can help develop better measures and methods more quickly. Evolutionary anthropologists, in particular, will be able to contribute much to our understanding and analysis of network structuring processes because they are explicitly interested in how things develop over time. This indicates that, at a minimum, evolutionary anthropologists can contribute considerable insight into these issues by working with network scientists who are developing measures and methods for the understanding and analysis of multilayer social networks.

Furthermore, we anticipate that theoreticians and methodologists will be interested in working with evolutionary anthropologists who study social networks because they oftentimes gather data which are inherently multiplex and will already have such data. As an example, many people evolutionary anthropologists work with practice mixed subsistence strategies. As such, they may have already asked questions such as "With whom do you hunt?", "With whom do you fish?", "With whom do you cut down fields?", and "With whom do you go to the market?" Therefore, if someone was working in a population that practices a mixed subsistence strategy, they may have already gathered multiplex networks in the course of gathering data on subsistence. Many other people interested in the effect of social networks on outcomes will have gathered similarly multiplex data, meaning that these datasets will be ideal for theoreticians and methodologists to work with.

Given the constraints of time and funding, it is unreasonable to suggest that everyone gather data on every single behavior or every single borrowed item (for instance). There are some quantitative measures that may help us decide which layers to use in an analysis once we have gathered the data (e.g., matrix correlation or information theoretic measures), but there are currently no methods that help us determine how many layers to gather before data collection. Therefore, it is incumbent on us to use our domain and ethnographic knowledge to think of the most salient layers that we can collect given our various constraints. It is doubtful that the patience of either the researcher or the respondent will be sufficient to gather all networks (CA's interviews in which participants could nominate alters on over 100 networks took hours to complete), but the logic underlying multiplex structuring processes leads us to believe that any attempt to gather more than one layer is better than none. As an example, the ongoing ENDOW project that is gathering complete network data from one or two communities in over 30 societies asks about six networks9. Once we have gathered data, algorithms implemented in programs such as Muxviz10 can help us decide how many layers to include in our final analyses.

In addition to being phenomena worthy of study in their own right, multiplex structuring processes complicate traditional network analysis. The structure of the multiplex can result in each layer being non-optimally arranged, giving an additional source of measurement error. We may gather data on a hunting network, for instance, and then try to predict some outcome, like frequency of hunting. If we fail to find an effect, that may be because we did not parse the hunting layer from the rest of the multiplex structure. It may be the case that, all things equal, being more central in a hunting layer leads to increased frequency of hunting, but it could also be the case that people central in the hunting network tend to be central in other layers, and these other networks prevent them from hunting at the frequency they might otherwise. The increased measurement error due to the structure of the multiplex network may mean that sometimes an effect of a layer is found, when it is actually due to a different part of the underlying structure (type 1 error). Sometimes it will mean that no effect of a single layer is found when there actually is an effect, but that might be because the effect of the underlying structure and the unique effect of that layer go in opposite directions, leading to a false detection of no effect (type 2 error). If the multiplex network structuring process is a common cause of measures on each layer (e.g., centrality), then we are unlikely to recover the true effect of each network on the outcome of interest unless we have a method to parse the centrality unique to each layer from the centrality in multiple layers. Given that we know processes such as incidental network membership lead to the coupling of networks, and that measures on coupled networks are not independent, we expect the creation of tools incorporating the structure of multiplex networks to be an active and productive area of research. As it stands, there exists no plug-and-play method for disentangling this coupling – we recommend working closely with experts in mechanistic statistical models.

The existence of these multiplex network structuring processes leads us to conclude that we must incorporate the structure of an individual's ties across domains or else we risk drawing wrong conclusions. In order to do this appropriately, we need to develop models and techniques for analyzing multiplexed settings. There has been relatively little work on multilayer and multiplex networks to date, but one of the areas to first receive attention is the concept of interdependence.3 In this context, interdependence means the likelihood of having ties in more than one layer.

How, then, should evolutionary anthropologists proceed given these considerations? The primary suggestion is to seek out theoreticians and methodologists to collaborate with: your combined insights will be much greater than could be arrived at independently. While those collaborations are being developed, we recommend gathering multiplex social network data as possible. Instead of simply asking, "Who do you borrow things from?", divide the question into a few things of interest such as food, cooking fuel, farm implements, and all else. Finally, even those who did not do so intentionally may have gathered social network data that is inherently multiplex, and they can consider how to separate their layers to treat them as a multiplex network. We are not recommending refraining from analyzing data that were collected as a single layer but, rather, interpreting those analyses with caution. There is much to be learned if kinship affects a behavior, even if we do not know exactly how kinship got translated to the outcome. However, any analysis of single-layer social networks may conflate the effect of that layer and the effect of the underlying multiplex structure. By tempering the interpretation of such analyses, we can guard against errors while our collaborations and new data collection are happening.

**6 Conclusion**
All humans are embedded in multiplex social networks: we have different partners in different domains of interaction. Multiplex structuring processes are ubiquitous. Specific categories of multiplex structuring processes that we have discussed here are based on different ways in which interdependencies between layers arise: similar influences on network formation of individual characteristics across layers, cross-domain dependencies, cross-domain complementarities, and spillover of interactions across domains.

We illustrated these processes with discussion of two specific examples. First, in incidental network membership a tie is formed between two individuals in a certain domain not because they are optimal partners for each other, but by virtue of them being connected (perhaps optimally) in another, more important layer. This illustrates the potential inefficiencies that may arise when one domain drives the formation of another. The second example we discussed, was a recent model based on coupling between layers of a multiplex network. An example of this sort of coupling across domains showed that it is possible to have large areas of the parameter space where network optimality may not be reached. These two examples show that multiplex structuring processes can lead to non-optimal networks, and that we should incorporate multiplex networks and their structuring processes into our analysis of the evolution of human behavior. While the development of techniques to incorporate these into our analysis is just beginning, there are already some promising directions and we expect that many more will be generated. Appreciating the multiplex and linked nature of the domains of interaction humans are involved in will not only add richness to our understanding but bring us to a better explanation for the causes of behavior.

Acknowledgements
We would like to acknowledge the Evolution and Ecology of Human Behavior and Culture group at University of California, Davis for comments. J.A.H. and P.G. were partially supported as RENOIR Project by the European Union Horizon 2020 research and innovation program under the Marie Skłodowska-Curie grant agreement No 691152 and by Ministry of Science and Higher Education (Poland), grant Nos. W34/H2020/2016, 329025/PnH /2016 and National Science Centre, Poland Grant No. 2015/19/B/ST6/02612. J.A.H. was partially supported by the Russian Scientific Foundation, Agreement #17-71-30029 with co-financing of Bank Saint Petersburg. M.O.J. acknowledges financial support under NSF grant SES-1629446. R.M.D. acknowledges financial support under U.S. Army Research Office MURI Award No. W911NF-13-1-0340.